\documentclass{epl}

\title{Confining strings in the Abelian-projected
$SU(3)$-gluodynamics II. 4D-case with $\theta$-term}
\author{D. Antonov\inst{1,2}}
\institute{
  \inst{1} INFN-Sezione di Pisa, Universit\'{a} degli studi di Pisa,
 Dipartimento di Fisica, Via Buonarroti, 
 2 - Ed. B - I-56127 Pisa, Italy\\
  \inst{2} Institute of Theoretical 
 and Experimental Physics, B. Cheremushkinskaya 25, 
 RU-117 218 Moscow, Russia}

\begin{document}

\maketitle

\begin{abstract}
The generalization of 4D confining 
string theory to the $SU(3)$-inspired case 
is derived. It 
describes string representation of the Wilson loop 
in the $SU(3)$-analogue of 
compact QED extended by the $\theta$-term.
It is shown that although
the obtained theory of confining strings 
differs from that of compact QED, their 
low-energy limits have the same functional form. This fact 
leads to the appearance of the string $\theta$-term
in the low-energy limit of the $SU(3)$-inspired 
confining string theory. In particular, it is shown that in the 
extreme strong coupling regime, the crumpling of 
string world sheets could disappear owing to the string 
$\theta$-term at $\theta=\pi/12$.
Finally, some characteristic features of the $SU(N)$-case 
are pointed out.
\end{abstract}

One of the ways to construct string representation of QCD is to use
the method of Abelian projections~\cite{abpr} (see ref.~\cite{rev}
for recent reviews). The assumption of monopole condensation 
employed within this method has recently been proved with a high
accuracy by the lattice calculations in ref.~\cite{novel}.
To model the condensation of Abelian-projected monopoles one can treat
them as a grand canonical ensemble with the Coulomb interaction.
In the 3D $SU(3)$-inspired case~\cite{2}, this led to a certain generalization
of the theory of confining strings~\cite{euro}, which was originally invented 
for the description of confinement of electric charges 
in 3D compact QED~\cite{confstr}. 
The aim of the present letter is to 
extend the results of ref.~\cite{euro} to the realistic 4D case 
with the inclusion of $\theta$-term.
The model in which confining strings will be studied
is thus nothing else, but the $SU(3)$-version of 
4D compact QED with $\theta$-term. In this way, we shall see 
that similarly to the case of compact QED~\cite{dia}, 
in the $SU(3)$-inspired model 
the field-theoretical $\theta$-term generates eventually string 
$\theta$-term. At a certain critical value of $\theta$, the latter one  
might be important for getting rid of crumpling of the string world sheet.

Consider first the case without the $\theta$-term.
Within the so-called Abelian dominance hypothesis~\cite{abdom}, 
the $SU(3)$-inspired theory under study 
describes free diagonal 
gluons and Abelian-projected monopoles ({\it cf.} the 
3D-case~\cite{2, snyd}). 
The action of this theory can be written as follows~\footnote{
Throughout the present letter, all the investigations will be 
performed in the Euclidean space-time.}:

\begin{equation}
\label{1}
\frac{1}{4}\int d^4x{\bf F}_{\mu\nu}^2+
\frac{1}{2}\int d^4x\int d^4y{\bf j}_\mu^{\rm mon}(x)
D_0(x-y){\bf j}_\mu^{\rm mon}(y).
\end{equation}
Here, 
${\bf F}_{\mu\nu}=\partial_\mu{\bf A}_\nu-\partial_\nu{\bf A}_\mu$
is the field strength tensor of diagonal gluons ${\bf A}_\mu=
\left(A_\mu^3, A_\mu^8\right)$ and $D_0(x)=1/(4\pi^2x^2)$ is the 4D 
massless propagator. Besides that, in eq.~(\ref{1}) 
the $N$-monopole current  
${\bf j}_\mu^{\rm mon}(x)$ is defined for $N>0$ as follows:

\begin{equation}
\label{2}
{\bf j}_\mu^{\rm mon}(x)=g_m\sum\limits_{a=1}^{N}{\bf q}_{\alpha_a}
\oint dz_\mu^a(\tau)\delta\left(x-x^a(\tau)\right),
\end{equation}
whereas for $N=0$, ${\bf j}_\mu^{\rm mon}(x)\equiv 0$.
Here, the magnetic coupling constant $g_m$ is related to the QCD  
coupling constant $g$ by the equation $gg_m=4\pi$. Next, 
in eq.~(\ref{2}), we have parametrized the trajectory of the 
$a$-th monopole by the vector $x_\mu^a(\tau)=y_\mu^a+z_\mu^a(\tau)$,
where $y_\mu^a=\int\limits_{0}^{1}d\tau x_\mu^a(\tau)$ is the 
position of the trajectory, whereas the vector $z_\mu^a(\tau)$ 
describes its shape, both of which should be averaged over. 
Finally, the magnetic charges of monopoles 
are described by six root vectors of the group $SU(3)$:
${\bf q}_1=(1/2, \sqrt{3}/2)$, ${\bf q}_2=(-1, 0)$, 
${\bf q}_3=(1/2, -\sqrt{3}/2)$, ${\bf q}_{-\alpha}=-{\bf q}_\alpha$.

The $\theta$-term by which we are going to extend the theory~(\ref{1})
originally having the form
$-\frac{i\theta g^2}{32\pi^2}\int d^4x{\bf G}_{\mu\nu}
\tilde{\bf G}_{\mu\nu}$ can be rewritten modulo full derivatives as 
$\frac{i\theta g^2}{8\pi^2}
\int d^4x{\bf A}_\mu{\bf j}_\mu^{\rm mon}$.
Here, $\tilde{\cal O}_{\mu\nu}=\frac12\varepsilon_{\mu\nu\lambda\rho}
{\cal O}_{\lambda\rho}$, 
${\bf G}_{\mu\nu}\equiv{\bf F}_{\mu\nu}+
{\bf F}_{\mu\nu}^{\rm mon}$,
and the monopole field strength tensor,
${\bf F}_{\mu\nu}^{\rm mon}$, is related to the 
current ${\bf j}_\mu^{\rm mon}$ according to the modified Bianchi identities:
$\partial_\mu\tilde{\bf F}_{\mu\nu}^{\rm mon}={\bf j}_\nu^{\rm mon}$.

Let us now concentrate ourselves on the derivation of an effective 
field theory describing the monopole ensemble. The desired monopole
part of the partition function, 
${\cal Z}_{\rm mon}\left[{\bf A}_\mu\right]$, (which should 
then be averaged {\it w.r.t.} the action 
$\frac{1}{4}\int d^4x{\bf F}_{\mu\nu}^2$) 
is nothing else,
but the statistical weight of the $N$-monopole configuration,

\begin{equation}
\label{aux}
{\cal Z}\left[{\bf j}_\mu^{\rm mon}, {\bf A}_\mu\right]=\exp\left[
-\frac{1}{2}\int d^4x\int d^4y{\bf j}_\mu^{\rm mon}(x)
D_0(x-y){\bf j}_\mu^{\rm mon}(y)-
\frac{i\theta g^2}{8\pi^2}
\int d^4x{\bf A}_\mu{\bf j}_\mu^{\rm mon}\right],
\end{equation}
summed up over the grand canonical ensemble of monopoles: 

\begin{equation}
\label{3}
{\cal Z}_{\rm mon}\left[{\bf A}_\mu\right]=
\sum\limits_{N=1}^{\infty}\frac{\zeta^N}{N!}\left<
{\cal Z}\left[{\bf j}_\mu^{\rm mon}, {\bf A}_\mu\right]
\right>.
\end{equation} 
Here, $\zeta\propto {\rm e}^{-S_{\rm mon}}$
is the fugacity (Boltzmann factor) of a single monopole loop, which has the 
dimension $({\rm mass})^4$. As far as the action of a single $a$-th 
loop is concerned, it obeys the  
estimate $S_{\rm mon}\propto\frac{1}{g^2}\int\limits_{0}^{1}d\tau
\sqrt{\left(\dot z_\mu^a\right)^2}$, where the integral expressing 
the length of the loop is supposed to be of the same order of magnitude
for all the loops. Next, the average over 
monopole loops in eq.~(\ref{3}) has the following form:

\begin{equation}
\label{4}
\left<{\cal O}\left[{\bf j}_\mu^{\rm mon}\right]\right>=
\prod\limits_{a=1}^{N}\int d^4y^a\int {\cal D}z^a\mu\left[z^a\right]
\sum\limits_{\alpha_a=\pm 1, \pm 2, \pm 3}^{}
{\cal O}\left[{\bf j}_\mu^{\rm mon}\right]
\end{equation}
In eq.~(\ref{4}), $\mu\left[z^a\right]$
stands for a certain rotation- and translation invariant measure of 
integration over the shapes of monopole loops. Similarly to
the case of compact QED~\cite{prepr}, the concrete form of this measure 
is irrelevant to the final result for the effective action.
It is only important that this measure 
is normalized in the standard way, $\left<1\right>=1$, and 
from now on this normalization condition will be implied.

Equation~(\ref{3}) can further be rewritten as follows:

$$
{\cal Z}_{\rm mon}\left[{\bf A}_\mu\right]=
\sum\limits_{N=1}^{\infty}\frac{\zeta^N}{N!}\int {\cal D}{\bf j}_\mu
{\cal Z}\left[{\bf j}_\mu, {\bf A}_\mu\right]\left<
\delta\left({\bf j}_\mu-{\bf j}_\mu^{\rm mon}\right)\right>=$$

\begin{equation}
\label{5}
=\int {\cal D}{\bf j}_\mu{\cal Z}\left[{\bf j}_\mu, {\bf A}_\mu\right]
\int {\cal D}{\bf l}_\mu\exp\left(-i\int d^4x{\bf l}_\mu
{\bf j}_\mu\right)\sum\limits_{N=1}^{\infty}\frac{\zeta^N}{N!}
\left<\exp\left(i\int d^4x{\bf l}_\mu
{\bf j}_\mu^{\rm mon}\right)\right>.
\end{equation}
The sum over $N$ here is equal to

$$\exp\left[2\zeta\int d^4y\sum\limits_{\alpha=1}^{3}
\int {\cal D}z\mu[z]\cos\left(g_m{\bf q}_\alpha\oint dz_\mu(\tau)
{\bf l}_\mu(x(\tau))\right)\right]\simeq$$

\begin{equation}
\label{6}
\simeq\exp\left[2\zeta\int d^4y\sum\limits_{\alpha=1}^{3}
\cos\left(\frac{{\bf q}_\alpha\left|{\bf l}_\mu(y)\right|}{\Lambda}
\right)\right].
\end{equation}
The derivation of this formula was based on a certain evaluation of 
the integral
over shapes of monopole loops in the strong coupling limit $g\gg 1$
analogous to that of compact QED~\cite{prepr}~\footnote{Note 
that similarly to
the case of 4D compact QED~\cite{dia, ell}, 
eq.~(\ref{6}) 
can also be derived by using the lattice regularization.}. 
Besides that, it has been 
assumed that 
characteristic distances between loops $\sqrt{\left(y_\mu^a\right)^2}$
as well as characteristic loop sizes $\sqrt{\left(z_\mu^a\right)^2}$ are
of the same order of magnitude for 
all $a$'s. The respective UV momentum cutoff $\Lambda$ 
(which is much larger than $1/\sqrt{\left(z_\mu^a\right)^2}$) 
obeys the estimate 
$\Lambda\sim g\sqrt{\left(y_\mu^a\right)^2}/\left(z_\mu^a\right)^2$. 
Finally, in eq.~(\ref{6}),
$\left|{\bf l}_\mu\right|$ denotes the modulus of the 
vector only {\it w.r.t.} the Lorentz indices, but not 
{\it w.r.t.} the Cartan ones, {\it i.e.} $\left|{\bf l}_\mu\right|\equiv
\left(\left|{\bf l}_\mu\right|^1,\left|{\bf l}_\mu\right|^2\right)=
\left(\sqrt{l_\mu^1l_\mu^1},\sqrt{l_\mu^2l_\mu^2}\right)$.

We should now carry out the integration over the Lagrange multiplier 
${\bf l}_\mu$ stemming from eqs.~(\ref{5}) and~(\ref{6}). 
Since the field ${\bf l}_\mu$ has no kinetic term, this can be 
done in the saddle-point approximation. For every Cartan index $a=1,2$, 
the respective saddle-point equation reads
$\frac{l_\mu^a}{\left|{\bf l}_\mu\right|^a}
\sum\limits_{\alpha=1}^{3}q_\alpha^a\sin
\left(\frac{{\bf q}_\alpha\left|{\bf l}_\mu\right|}{\Lambda}\right)=
-\frac{i\Lambda}{2\zeta}j_\mu^a$. 
It can be solved {\it w.r.t.} ${\bf q}_\alpha\left|{\bf l}_\mu\right|$
by adapting for $l_\mu^a$ the following {\it Ansatz}
$l_\mu^a=
\left|{\bf l}_\mu\right|^aj_\mu^a/\left|{\bf j}_\mu\right|^a$
and representing $\left|{\bf j}_\mu\right|^a$ 
as $\sum\limits_{\alpha=1}^{3}q_\alpha^a
j^{(\alpha)}$, where $j^{(1)}=\left(\left|{\bf j}_\mu\right|^1/
\sqrt{3}+\left|{\bf j}_\mu\right|^2
\right)/\sqrt{3}$, $j^{(2)}=-2\left|{\bf j}_\mu\right|^1/3$, $j^{(3)}=
\left(\left|{\bf j}_\mu\right|^1/\sqrt{3}-\left|{\bf j}_\mu\right|^2
\right)/\sqrt{3}$. 
The so-obtained solution reads ${\bf q}_\alpha\left|{\bf l}_\mu\right|=
-i\Lambda~ {\rm arcsinh}\left(\Lambda j^{(\alpha)}/
(2\zeta)\right)$. By making use of it and 
performing the rescaling of fields $g{\bf A}_\mu=
{\bf A}_\mu^{\rm new}$, $g{\bf j}_\mu={\bf j}_\mu^{\rm new}$ 
we eventually get from eqs.~(\ref{1}), (\ref{aux}), and (\ref{5})
the following expression for the full 
partition function: ${\cal Z}=\int {\cal D}{\bf A}_\mu{\cal D}{\bf j}_\mu
{\rm e}^{-S}$, where 

\begin{equation}
\label{s}
S=\frac{1}{4g^2}\int d^4x{\bf F}_{\mu\nu}^2+\frac{1}{2g^2}
\int d^4x\int d^4y
{\bf j}_\mu(x)D_0(x-y){\bf j}_\mu(y)+\frac{i\theta}{8\pi^2}
\int d^4x {\bf A}_\mu{\bf j}_\mu+V\left[{\bf j}_\mu\right].
\end{equation}
Here, 

\begin{equation}
\label{7}
V\left[{\bf j}_\mu\right]=\sum\limits_{\alpha=1}^{3}\int d^4x\left[
\frac{\Lambda}{g} j^{(\alpha)}{\,} {\rm arcsinh}\left(\frac{\Lambda}{2g\zeta}
j^{(\alpha)}\right)-2\zeta\sqrt{1+
\left(\frac{\Lambda}{2g\zeta}
j^{(\alpha)}\right)^2}\right]
\end{equation}
is the multivalued potential of monopole currents.

As we shall see below, it is convenient to unify the 
kinetic term of the field ${\bf A}_\mu$ and the interaction 
of monopole currents by introducing the following antisymmetric 
tensor field (else called Kalb-Ramond 
field~\cite{kr})~\footnote{In the formal language, this equation represents
the Hodge decomposition theorem.} 
${\bf B}_{\mu\nu}={\bf F}_{\mu\nu}+{\bf h}_{\mu\nu}$. Here, the field 
${\bf h}_{\mu\nu}$ is unambiguously related to the current ${\bf j}_\mu$ 
as ${\bf h}_{\mu\nu}=-\varepsilon_{\mu\nu\lambda\rho}\partial_\lambda^x
\int d^4yD_0(x-y){\bf j}_\rho(y)$ and thus obeys the modified Bianchi 
identities $\partial_\mu \tilde{\bf h}_{\mu\nu}={\bf j}_\nu$.~\footnote{
Clearly, the same Legendre transformation, which made from the current
${\bf j}_\mu^{\rm mon}$ the dynamical field ${\bf j}_\mu$, makes from 
${\bf F}_{\mu\nu}^{\rm mon}$ the dynamical field ${\bf h}_{\mu\nu}$.}
In terms of the field ${\bf B}_{\mu\nu}$, the action~(\ref{s}) takes the 
following more compact form:

\begin{equation}
\label{8}
S=\frac{1}{4g^2}\int d^4x{\bf B}_{\mu\nu}^2-\frac{i\theta}{32\pi^2}
\int d^4x{\bf B}_{\mu\nu}\tilde{\bf B}_{\mu\nu}+V\left[\partial_\mu
\tilde{\bf B}_{\mu\nu}\right].
\end{equation}

In what follows, we shall be interested in the string representation 
of the Wilson loop defined as 
$\left<W(C)\right>=\frac13\left<{\rm tr}{\,}
P{\,}\exp\left(i\oint
\limits_{C}^{}dx_\mu{\bf A}_\mu^{\rm tot}{\bf T}\right)
\right>$.
Here, ${\bf A}_\mu^{\rm tot}$ is the total vector potential which includes
also the monopole contributions, and 
${\bf T}=\left(\frac{\lambda_3}{2},\frac{\lambda_8}{2}\right)$
with $\lambda_{3,8}$ denoting the Gell-Mann matrices. Analogously to the 
case when monopoles are absent, Stokes theorem yields for 
the Wilson loop the following expression:

\begin{equation}
\label{w}
\left<W(C)\right>=\frac13\left<
{\rm tr}{\,}\exp\left[\frac{i}{2}
\int d^4x{\bf G}_{\mu\nu}
{\bf T}\Sigma_{\mu\nu}\right]\right>
\end{equation}
Here, 
$\Sigma_{\mu\nu}[x, \Sigma]=\int\limits_{\Sigma(C)}^{} 
d\sigma_{\mu\nu}(x(\xi))\delta(x-x(\xi))$
is the vorticity tensor current defined at a certain surface 
$\Sigma(C)$ bounded by the contour $C$ and parametrized by the vector 
$x_\mu(\xi)$ with $\xi=(\xi^1,\xi^2)$ standing for the 2D coordinate.
In a derivation of eq.~(\ref{w}), 
we have omitted the path ordering, which 
is possible due to the fact that both $\lambda_3$ and $\lambda_8$ are
diagonal. Note that by virtue of the quantization condition 
$gg_m=4\pi$ one can prove that 
eq.~(\ref{w}) is indeed independent 
of the form of the surface $\Sigma$. 

The average in eq.~(\ref{w})
is first performed over the free part of the ${\bf A}_\mu^{\rm tot}$-action,
$\frac{1}{4g^2}\int d^4x{\bf F}_{\mu\nu}^2$, 
after which the result should be weighted with 
${\cal Z}\left[\frac1g{\bf j}_\mu^{\rm mon}, \frac1g{\bf A}_\mu\right]$ 
and summed up over the grand canonical ensemble of 
monopoles in the sense of eqs.~(\ref{3})-(\ref{4}). 
This procedure can be simplified by rewriting the Wilson loop
in terms of the above-introduced field ${\bf B}_{\mu\nu}$ as follows:

\begin{equation}
\label{9}
\left<W(C)\right>
=\frac13\sum\limits_{\alpha=1}^{3}\left<
\exp\left(\frac{i}{2}\int d^4x{\bf B}_{\mu\nu}{\bf Q}_\alpha
\Sigma_{\mu\nu}\right)\right>.
\end{equation}
Here, $\left<\ldots\right>$ stands for the average {\it w.r.t.} the 
action~(\ref{8}), and the vectors 
${\bf Q}_\alpha$'s, which denote the charges of
a quark of the $\alpha$'s colour {\it w.r.t.} the diagonal gluons,
have the following form:
${\bf Q}_1=\left(
-\frac12,\frac{1}{2\sqrt{3}}\right)$, ${\bf Q}_2=
\left(\frac12,\frac{1}{2\sqrt{3}}\right)$, ${\bf Q}_3=
\left(0,-\frac{1}{\sqrt{3}}\right)$.  
Besides that, in the 
derivation of eq.~(\ref{9}) we have used 
the identity 
${\rm tr}{\,}\exp\left(i{\bf a}{\bf T}\right)=
\sum\limits_{\alpha=1}^{3}\exp\left(i{\bf a}{\bf Q}_\alpha\right)$
valid for an arbitrary vector ${\bf a}$.

In the low-energy limit, 
$\left|j^{(\alpha)}\right|\ll g\zeta/\Lambda$, 
the monopole potential~(\ref{7})  
becomes a quadratic functional. This yields the kinetic term of the field 
${\bf B}_{\mu\nu}$,
$V\left[\partial_\mu
\tilde{\bf B}_{\mu\nu}\right]\to\frac{1}{12\eta^2}\int d^4x
{\bf H}_{\mu\nu\lambda}^2$,
where $\eta\equiv g\sqrt{3\zeta}/\Lambda$ 
and ${\bf H}_{\mu\nu\lambda}=\partial_\mu {\bf B}_{\nu\lambda}+
\partial_\lambda {\bf B}_{\mu\nu}+\partial_\nu {\bf B}_{\lambda\mu}$.
The Wilson loop~(\ref{9}) in such a low-energy limit 
takes the form

$$\left.\left<W(C)\right>\right|_{\rm low{\,}en.}=
\frac{1}{{\cal Z}_{\rm low{\,}en.}}\times$$

\begin{equation}
\label{10}
\times
\frac13\sum\limits_{\alpha=1}^{3}\int {\cal D}{\bf B}_{\mu\nu}
\exp\left\{-\int d^4x\left[
\frac{1}{12\eta^2}{\bf H}_{\mu\nu\lambda}^2+
\frac{1}{4g^2}{\bf B}_{\mu\nu}^2-\frac{i\theta}{32\pi^2}
{\bf B}_{\mu\nu}\tilde{\bf B}_{\mu\nu}-\frac{i}{2}
{\bf B}_{\mu\nu}{\bf Q}_\alpha
\Sigma_{\mu\nu}\right]\right\},
\end{equation}
where ${\cal Z}_{\rm low{\,}en.}$ is given by the second line 
of eq.~(\ref{10}) with $\Sigma_{\mu\nu}$ set to zero.
Carrying out the ${\bf B}_{\mu\nu}$-integration, whose details 
are outlined in the Appendix A, we get the following result:

$$\left.\left<W(C)\right>\right|_{\rm low{\,} en.}=
\exp\left\{-\frac{1}{12}\left[2g^2\oint\limits_{C}^{}dx_\mu
\oint\limits_{C}^{}dy_\mu D_m(x-y)+\right.\right.$$

\begin{equation}
\label{11}
\left.\left.+\eta^2\int d^4x\int d^4y
D_m(x-y)\left(\Sigma_{\mu\nu}(x)\Sigma_{\mu\nu}(y)+
\frac{i\theta g^2}{8\pi^2}\Sigma_{\mu\nu}(x)\tilde\Sigma_{\mu\nu}(y)
\right)\right]\right\}.
\end{equation}
Here, 

\begin{equation}
\label{mass}
m=\frac{\eta}{g}\sqrt{1+\left(\frac{\theta g^2}{8\pi^2}\right)^2}
\end{equation}
is the mass of the field ${\bf B}_{\mu\nu}$, and 
$D_m(x)=mK_1(m|x|)/\left(4\pi^2|x|\right)$ is the massive boson 
propagator with $K_1$ standing for the modified Bessel function. 

Upon the derivative expansion 
of eq.~(\ref{11}) (analogous to the  
expansion performed in ref.~\cite{deriv} 
within the stochastic vacuum model of QCD),  
one gets as a few first local string terms the usual Nambu-Goto and rigidity
terms, responsible for confinement of electric charges 
and stability of strings. Besides that, this expansion yields  
string $\theta$-term~\cite{teta}~\footnote{ 
The string $\theta$-term can also be derived from the instanton gas model
of QCD, as it has been done in ref.~\cite{prd}.} equal to $ic\nu$, where
$\nu\equiv (2\pi)^{-1}\int d^2\xi\sqrt{\hat g}\hat g^{ab}\left(\partial_a
t_{\mu\nu}\right)\left(\partial_b\tilde t_{\mu\nu}\right)$ is 
the number of self-intersections of the world sheet 
$\Sigma$. Here we have adapted the standard notations:  
$\hat g^{ab}=
(\partial^ax_\mu(\xi))(\partial^bx_\mu(\xi))$
denotes the induced metric tensor of the surface, $\hat g=\det\left|\left|
\hat g^{ab}\right|\right|$, and 
$t_{\mu\nu}=\varepsilon^{ab}
\left(\partial_ax_\mu(\xi)\right)\left(\partial_bx_\nu(\xi)\right)/
\sqrt{\hat g}$ stands for its extrinsic curvature tensor. 
The coupling constant $c$ can be calculated analogously to 
refs.~\cite{deriv, prd} and reads $c=-\frac{\theta}{768}\left(
\frac{g\eta}{\pi m}\right)^2$. Since the obtained string $\theta$-term 
is proportional to the number of self-intersections of the world sheet,
it might be relevant to the 
solution of the problem of crumpling of the world sheet, as it has 
been first mentioned in ref.~\cite{teta}. In particular, in the 
extreme strong coupling limit $g\to\infty$, $c\to-\frac{\pi^2}{12\theta}$,
and therefore self-intersections are weighted in the partition function 
with the desired factor $(-1)^{\nu}$ at $\theta=\pi/12$. 

Notice also that
owing to the field-theoretical $\theta$-term, in this limit the 
dual mass~(\ref{mass}) is very large: 

\begin{equation}
\label{large}
m\to\frac{\theta g\eta}{8\pi^2}.
\end{equation}
This situation is opposite to the case without $\theta$-term, 
where the dual mass
equal to $\eta/g$ vanishes in the extreme strong coupling limit, since 
$\eta(g)\propto {\rm e}^{-{\rm const}/g^2}\to 1$. 
Therefore, contrary to that case, at $\theta\ne 0$, the expansion of the 
$\Sigma$-dependent part of the action~(\ref{11}) in powers
of the derivatives {\it w.r.t.} $\xi^a$'s~\cite{deriv, prd}, being 
nothing else but the expansion in the inverse powers of $m$, converges fastly.
In particular, this means that all the terms of this expansion higher  
in the derivatives than the Nambu-Goto, rigidity, and $\theta$-ones
are irrelevant, since they are suppressed by the higher powers of the mass.

Clearly, 
all the lowest local string terms, {\it i.e.} the Nambu-Goto and rigidity
ones, as well as the above-discussed $\theta$-term, depend 
explicitly on the form of the world sheet $\Sigma$. This is 
the consequence of the fact that in a derivation of eq.~(\ref{10})
we have taken into account only one (namely, real) branch of the 
monopole potential~(\ref{7}). However, as it has been argued
first for the case of compact QED in ref.~\cite{confstr}
and then discussed for the 3D $SU(3)$-inspired case in ref.~\cite{euro},
the world-sheet independence of the Wilson loop  
becomes restored upon the 
summation over all the complex-valued branches of the 
potential~(\ref{7}) at every space-time point. 

Notice that although 
the low-energy limits of confining string theories obtained 
in compact QED and in the $SU(3)$-inspired case under study
have the same functional form, the full expressions  
are different since in the $SU(3)$-case the ${\bf B}_{\mu\nu}$-field 
has two components. Moreover,  
the obtained results can 
be generalized to the $SU(N)$-inspired case along with the 
lines of ref.~\cite{snyd}. 
There the fields 
${\bf B}_{\mu\nu}$ and ${\bf j}_\mu$ have 
$(N-1)$ components, but the functional form of the action~(\ref{s})-(\ref{7})
remains the same modulo the fact that the fields $j^{(\alpha)}$'s are given
by more complicated combinations of the fields 
$\left|{\bf j}_\mu\right|^a$'s. Let us finally discuss the 
asymptotic behaviour $m(N)$ at $N\gg 1$. Note first of all that the
strong coupling limit under consideration is still accessible at 
large $N$ despite the fact that $g\sim 1/\sqrt{N}$. That is because
the strong coupling limit implies only that $g$ should be
larger than some critical value, which itself scales as $1/\sqrt{N}$.
Taking into account that $\eta(g)\propto {\rm e}^{-{\rm const}/g^2}$, 
we see that according to eq.~(\ref{mass}), the dual mass behaves in the 
large-$N$ limit as $\sqrt{N}{\rm e}^{-{\rm const}{\,}N}$.
Besides that, in the $SU(N)$-case there exist $N(N-1)/2$ vectors 
${\bf q}_\alpha$'s with $\alpha>0$ and consequently the same amount
of mass terms of $(N-1)$ fields, which emerge  
from the expansion of cosines 
in the respective sine-Gordon theory describing these fields.
Therefore, the number of mass terms of every field is of the order of 
$N$, {\it i.e.} each mass scales as $\sqrt{N}$ owing to this fact. 
These two observations considered together 
lead to the conclusion that at $N\gg 1$, $m\sim N
{\rm e}^{-{\rm const}{\,}N}$.

However, in the large-$N$ limit, the 
model under study becomes less and less relevant to QCD.
That is because the number of off-diagonal fields disregarded
within the Abelian dominance hypothesis, 
equal to $\left(N^2-N\right)$, significantly exceeds in this limit
the number of diagonal fields kept, equal to $(N-1)$. The problem of 
accounting for off-diagonal degrees of freedom deserves special  
investigations and will be considered in future publications.

\acknowledgments
The author is indebted to Prof.  
A. Di Giacomo for helpful discussions.
He also acknowledges discussions with Profs.
A.E. Dorokhov and Yu.A. Simonov 
and Drs. N.O. Agasian and 
E. Meggiolaro. Besides that, the 
author is greatful to Prof. A. Di Giacomo and 
the whole staff of the Quantum Field Theory Division
of the University of Pisa for cordial hospitality and to INFN for the  
financial support.

\section{Appendix A. Details of integration over the ${\bf B}_{\mu\nu}$-field 
in eq.~(\ref{10})}
The desired integration is Gaussian and can obviously be performed
by finding the inverse to the quadratic part of the 
${\bf B}_{\mu\nu}$-field action. Since the integral under study 
factorizes into the product of the integrals over $B_{\mu\nu}^a$'s,
we should only solve the following equation in the momentum representation:

$$
\frac12\left(\frac{p^2}{\eta^2}\hat P+\frac{1}{g^2}\hat 1-
\frac{i\theta}{16\pi^2}\varepsilon\right)_{\mu\nu\alpha\beta}
G_{\alpha\beta\lambda\rho}(p)=\hat 1_{\mu\nu\lambda\rho}.\eqno(A.1)$$
Here, the following projection operators were introduced 
(see {\it e.g.}~\cite{prepr}): 

$$\hat P_{\mu\nu\alpha\beta}=\frac12
\left({\cal P}_{\mu\alpha}{\cal P}_{\nu\beta}-{\cal P}_{\mu\beta}
{\cal P}_{\nu\alpha}\right),~ \hat 1_{\mu\nu\alpha\beta}=\frac12
\left(\delta_{\mu\alpha}\delta_{\nu\beta}-\delta_{\mu\beta}
\delta_{\nu\alpha}\right),$$ 
where ${\cal P}_{\mu\nu}=\delta_{\mu\nu}-p_\mu p_\nu/p^2$.
In what follows, we shall for brevity omit the indices and 
use the notation $\left({\cal O}\cdot{\cal O}'\right)_{\mu\nu\lambda\rho}
\equiv{\cal O}_{\mu\nu\alpha\beta}{\cal O}'_{\alpha\beta\lambda\rho}$.
Then, representing $G$ as $G^{(1)}+iG^{(2)}$, we get the following 
equations for the real and imaginary parts of eq.~(A.1):

$$
\left(\frac{p^2}{\eta^2}\hat P+\frac{1}{g^2}\hat 1
\right)\cdot G^{(1)}+
\frac{\theta}{16\pi^2}\varepsilon\cdot
G^{(2)}=2
\hat 1,~ 
\left(\frac{p^2}{\eta^2}\hat P+\frac{1}{g^2}\hat 1
\right)\cdot G^{(2)}-
\frac{\theta}{16\pi^2}\varepsilon\cdot
G^{(1)}=0.$$
Applying further to the second of these equations the operation 
$\varepsilon\cdot$, it is possible to resolve it 
{\it w.r.t.} $G^{(1)}$. Substituting the so-obtained expression 
into the first equation, we get the following equation for $G^{(2)}$:

$$
\left(\frac{p^2}{\eta^2}+\frac{1}{g^2}\right)\left\{\frac{p^2}{\eta^2}
\left[\left(\hat P-\hat 1\right)\cdot\varepsilon+\varepsilon\cdot
\left(\hat P-\hat 1\right)\right]+\left(\frac{p^2}{\eta^2}
+\frac{1}{g^2}\right)\varepsilon\right\}\cdot G^{(2)}+
\frac{\theta^2}{64\pi^4}\varepsilon\cdot G^{(2)}=
\frac{\theta}{2\pi^2}\hat 1.$$
Finally, seeking for $G^{(2)}$ in the form of the {\it Ansatz} 
$G^{(2)}=f(p)\varepsilon$ and using the equation 
$\left[\left(\hat P-\hat 1\right)\cdot\varepsilon+\varepsilon\cdot
\left(\hat P-\hat 1\right)\right]\cdot\varepsilon=-4\hat 1$,
we find  

$$f=\frac{\theta g^2\eta^2}{8\pi^2\left(p^2+m^2\right)},~ 
G^{(1)}=\frac{2(g\eta)^2}{p^2+m^2}\left[\frac{p^2}{\eta^2}
\left(\hat 1-\hat P\right)+\frac{1}{g^2}\hat 1\right],$$
where the mass $m$ of the field ${\bf B}_{\mu\nu}$ is given by 
eq.~(\ref{mass}). One can now straightforwardly obtain  
eq.~(\ref{11}) of the main text by passing back to the 
coordinate representation, using Stokes theorem, and the 
fact that for every $\alpha$, ${\bf Q}_\alpha^2=1/3$.


\newpage

\begin{thebibliography}{100}

\bibitem{abpr}
\Name{'t Hooft G.} 
 \REVIEW{Nucl. Phys. B}{190}{1981}{455}.

\bibitem{rev}
 \Name{Di Giacomo A.}
 \REVIEW{Nucl. Phys. A}{661}{1999}{13}; 
 \REVIEW{Nucl. Phys. A}{663-664}{2000}{199}; 
 preprint hep-lat/9912016 (1999);
 preprint hep-lat/0012013 (2000).

\bibitem{novel}
\Name{Di Giacomo A., Lucini B., Montesi L., and Paffuti G.}
 \REVIEW{Phys. Rev. D}{61}{2000}{034503}; 
 \REVIEW{Phys. Rev. D}{61}{2000}{034504}.

\bibitem{2}
\Name{Wadia S.R., Das S.R.}
\REVIEW{Phys. Lett. B}{106}{1981}{386}.

\bibitem{euro}
\Name{Antonov D.}
\REVIEW{Europhys. Lett.}{52}{2000}{54}.

\bibitem{confstr}
\Name{Polyakov A.M.}
\REVIEW{Nucl. Phys. B}{486}{1997}{23}.

\bibitem{dia}
\Name{Diamantini M.C., Quevedo F., and Trugenberger C.A.}
\REVIEW{Phys. Lett. B}{396}{1997}{115}.

\bibitem{abdom}
\Name{Ezawa Z.F., Iwazaki A.}
\REVIEW{Phys. Rev. D}{25}{1982}{2681}; 
\REVIEW{Phys. Rev. D}{26}{1982}{631}.


\bibitem{snyd}
\Name{Snyderman N.J.}
\REVIEW{Nucl. Phys. B}{218}{1983}{381}.  


\bibitem{prepr}
\Name{Antonov D.}
\REVIEW{Surveys High Energy Phys.}{14}{2000}{265}.

\bibitem{ell}
\Name{Ellwanger U.}
\REVIEW{Nucl. Phys. B}{560}{1999}{587}.


\bibitem{kr}
\Name{Kalb M., Ramond P.} 
\REVIEW{Phys. Rev. D}{9}{1974}{2273}.

\bibitem{deriv}
\Name{Antonov D.V., Ebert D., and Simonov Yu.A.}
\REVIEW{Mod. Phys. Lett. A}{11}{1996}{1905}.

\bibitem{teta}
\Name{Polyakov A.M.}
\REVIEW{Nucl. Phys. B}{268}{1986}{406}.

\bibitem{prd}
\Name{Antonov D.V., Ebert D.}
\REVIEW{Phys. Rev. D}{58}{1998}{067901}.



\end{thebibliography}
\end{document}